\documentclass[mathleft
]{an}
\usepackage{graphicx}
\usepackage{times}
\usepackage{natbib}
\bibpunct{(}{)}{;}{a}{}{,}
\usepackage{hyperref}
\usepackage[space]{grffile}
\usepackage[]{amsmath}
\usepackage[normalem]{ulem}
\usepackage{comment}

\newcommand{\fek}{Fe~K$\alpha$}
\newcommand{\xmm}{{\em XMM--Newton}}
\newcommand{\nustar}{{\em NuSTAR}}

\newcommand{\suzaku}{{\em Suzaku}}
\newcommand{\sax}{{\em BeppoSAX}}
\newcommand{\integral}{{\em INTEGRAL}}
\newcommand{\cc}{NGC~5548}
\newcommand{\qc}{NGC~4593}

\defcitealias{kaastra2014science}{K14}
\defcitealias{5548}{U15}
\defcitealias{prova}{M14}
\begin{document}

\Pagespan{1}{}
\Yearpublication{}
\Yearsubmission{}
\Month{}
\Volume{}  
\Issue{}

\title{High-energy monitoring of Seyfert galaxies: the case of NGC~4593
	}

\author{F. Ursini\inst{1,2,3}\fnmsep\thanks{Corresponding author:
  \email{francesco.ursini@obs.ujf-grenoble.fr}}
\and P.-O. Petrucci\inst{1,2} \and G. Matt\inst{3} \and S. Bianchi\inst{3} \and M. Cappi\inst{4} \and B. De Marco\inst{5} \and A. De Rosa\inst{6} \and J.~Malzac\inst{7,8} \and G. Ponti\inst{5}
}
\titlerunning{High-energy monitoring of NGC 5548 and NGC 4593}
\authorrunning{F. Ursini et al.}
\institute{
Univ. Grenoble Alpes, IPAG, F-38000 Grenoble, France
\and 
CNRS, IPAG, F-38000 Grenoble, France
\and 
Dipartimento di Matematica e Fisica, Universit\`a degli Studi Roma Tre, via della Vasca Navale 84, 00146 Roma, Italy
\and 
INAF-IASF Bologna, Via Gobetti 101, I-40129 Bologna, Italy 
\and 
Max-Planck-Institut f\"ur extraterrestrische Physik, Giessenbachstrasse, D-85748 Garching, Germany 
\and
INAF/Istituto di Astrofisica e Planetologia Spaziali, via Fosso del Cavaliere, 00133 Roma, Italy
\and 
Universit\'e de Toulouse, UPS-OMP, IRAP, Toulouse, France
\and
CNRS, IRAP, 9 Av. colonel Roche, BP44346, F-31028 Toulouse cedex 4, France
}
\received{}
\accepted{}
\publonline{}


\abstract{%
  We discuss preliminary results from a joint \xmm~and \nustar~monitoring program on the active galactic nucleus \qc, consisting of $5 \times 20$ ks observations, spaced by two days, performed in January 2015. The source is found to be variable, both in flux and spectral shape, on time scales as short as a few ks. The spectrum clearly softens when the source brightens. A simple timing analysis suggests the presence of a variable soft excess that correlates with the primary continuum. 
  }

\maketitle

\section{Introduction}\label{sec:intro}
The central engine of active galactic nuclei (AGNs) is thought to be powered by accretion of surrounding matter onto a supermassive black hole. According to the standard paradigm, the accretion disc is responsible for the optical/UV emission, while the most likely origin of the X-ray emission is thermal Comptonization of the soft photons emitted by the disc in a hot region, the so-called corona \citep[see, e.g.,][]{haardt&maraschi1991,hmg1994,hmg1997}. This process is naturally able to explain the power-law shape of the observed X-ray spectrum of AGNs. A signature of thermal Comptonization is the presence of a high-energy cut-off, which is found in a number of cases \citep[see, e.g.,][]{perola2002,2041-8205-782-2-L25,IC4329A_Brenneman,marinucci2014swift,ballantyne2014,balokovic2015,5548}. The primary X-ray emission can be modified by diverse processes, such as absorption from neutral or ionized gas, and Compton reflection from the disc or more distant material, like the putative obscuring torus. Moreover, a smooth rise below 1-2 keV above the extrapolated high-energy power law is commonly observed in the spectra of AGNs \citep[see, e.g.,][]{caixa1}. The origin of this so-called soft excess is still uncertain \citep[see, e.g.,][]{done2012SE}. Ionized reflection provides a good explanation in a number of sources \citep[see, e.g.,][]{walton2013}, while a warm Comptonization mechanism is favoured in other cases \citep[see, e.g.,][]{rozenn2014mrk509SE}. \\ \\
The current understanding of the geometrical and physical properties of the X-ray corona is far from being complete. Multiple, broad-band observations with a high signal-to-noise ratio are needed to disentangle the different spectral components through the analysis of their variability and constrain their characteristic parameters. This approach has proven successful to study the origin of the high-energy emission in Seyfert galaxies, as shown by recent campaigns on Mrk~509 \citep{kaastra2011mrk509,pop2013mrk509}  and \cc~\citep{kaastra2014science,5548}. In particular \nustar, thanks to its dramatically improved signal-to-noise up to $\sim 80$ keV, allows for the study of the high-energy emission of AGNs with high precision (see, e.g., the review by Marinucci et al., presented in these proceedings). \\ \\
\cc~was the object of a long, multiwavelength campaign conducted from May 2013 to February 2014, with the main goal of studying its persistent ionized outflow, the so-called warm absorber \citep[][]{kaastra2014science}. During the campaign, the nucleus was surprisingly found to be obscured by a fast, massive and clumpy stream of gas never seen before in this source. The properties of this obscuring outflow are consistent with those of a disc wind \citep{kaastra2014science}. The detailed description of the campaign is given by \cite{mehdipour}, while the physical and temporal properties of the outflows are described in \cite{arav} and \cite{digesu}. Using the high-energy \xmm, \nustar~and \integral~data from the campaign, \citet[][]{5548} studied the broad-band (0.3--400 keV) X-ray spectrum of \cc. A high-energy cut-off of $70^{+40}_{-10}$ keV, attributed to thermal Comptonization, is found in one observation out of seven, while a lower limit ranging from 50 to 250 keV is found in the other six observations. The reflection component is consistent with being constant, thus it is likely produced by material lying a few light months away from the primary source. Furthermore, the average spectrum is well fitted by a Comptonized component produced by a corona with a temperature $kT_{\textrm{e}} = 40^{+40}_{-10}$ keV and an optical depth $ \tau =2.7^{+0.7}_{-1.2}$, assuming a spherical geometry. \\ \\
The long-term campaign on \cc~provided a detailed ``anatomy'' of that AGN. However, multiple observations on shorter time scales are equally important to make progress in the physical interpretation of the high-energy emission of AGNs. In the following, we discuss preliminary results from a monitoring program on the AGN \qc~with \xmm~and \nustar. We focus on a phenomenological timing analysis, while the spectral analysis will be presented in a forthcoming paper. In Sec. \ref{sec:4593} we describe the observations and an analysis of the temporal properties of the source. In Sec. \ref{sec:discuss}, we summarize our results and conclusions.

\section{The high-energy view of \qc}\label{sec:4593}
 \qc~\citep[$z=0.009$,][]{z4593} is an X-ray bright Seyfert 1 galaxy, hosting a supermassive black hole of $(9.8 \pm 2.1) \times 10^6$ solar masses \citep{mbh4593}. Previous observations of this source with \sax~ \citep{guainazzi1999}, \xmm~\citep{reynolds2004,brenneman2007} and \suzaku~\citep{markowitz2009} have shown a strong reflection hump above 10 keV \citep{guainazzi1999,markowitz2009} and the presence of two narrow emission features due to neutral and hydrogen-like Fe K$\alpha$ lines at 6.4 and 6.97 keV respectively \citep{reynolds2004,brenneman2007,markowitz2009}. \citet{reynolds2004} found the full width at half-maximum (FWHM) of the neutral Fe K$\alpha$ line to be $\sim 11000$ km s$^{-1}$ using a 2002 \xmm~observation, significantly larger than the FWHM of the H$\beta$ line, i.e. $4900 \pm 300$ km s$^{-1}$ \citep{grupe2004}. This result seems to suggest that the \fek~line-emitting material lies significantly inside the broad-line region (BLR). However, \citet{markowitz2009} found the FWHM of the \fek~line to be $\sim 4000$ km s$^{-1}$ in a 2007 \suzaku~observation, suggesting emission from at least $5000$ gravitational radii. The \fek~line-emitting region seems therefore to be changing with time. A soft excess below 2 keV was found both in the 2002 \xmm~data \citep{brenneman2007} and in the 2007 \suzaku~data, albeit with a drop in the 0.4--2 keV flux by a factor $>20$ in the latter \citep{markowitz2009}. Finally, \cite{guainazzi1999} found only a lower limit on the high-energy cut-off of 150 keV using \sax~data. 

\subsection{Observations and data reduction}\label{subsec:obs}
\xmm~\citep{xmm} and \nustar~\citep{harrison2013nustar} performed five joint observations of \qc, $\sim 20$ ks each, spaced by two days, starting from 2014 December 29. The log of the data sets is reported in Table \ref{log}.\\ \\
In the \xmm~observations, the EPIC instruments were operating in the Small-Window mode with the thin-filter applied. For simplicity, we used the EPIC-pn data only for our analysis. The data were processed using the \xmm~Science Analysis System (\textsc{sas} v13). Source extraction radii and screening for high-background intervals were performed through an iterative process which leads to a maximization of the signal-to-noise ratio, as described in \citet{pico2004}. 
\\ \\ 
The \nustar~data were reduced with the standard pipeline (\textsc{nupipeline}) in the \nustar~Data Analysis Software (\textsc{nustardas}, v1.3.1; part of the \textsc{heasoft} distribution as of version 6.14), using calibration files from \nustar~ {\sc caldb} v20150316. Spectra and light curves were extracted from the cleaned event files using the standard tool {\sc nuproducts} for each of the two hard X-ray detectors aboard \nustar, which sit inside the corresponding focal plane modules A and B (FPMA and FPMB). 
The source data were extracted from circular regions with a radius of 75 arcsec, and background was extracted from a blank area close to the source. 
\begin{table}
	\begin{center}
		\scriptsize
		\caption{The logs of the joint \textit{XMM-Newton} and \textit{NuSTAR} observations of NGC 4593. \label{log}}
		\begin{tabular}{ c c c c c } 
			\hline \rule{0pt}{2.5ex} Obs. &  Satellites & Obs. Id. & Start time (UTC)  & Net exp.\\ 
			& & & yyyy-mm-dd & (ks)  \\ \hline \rule{0pt}{2.5ex}
			1 & \xmm & 0740920201&  2014-12-29  & 16\\ 
			& \nustar & 60001149002 &  &  22\\ \hline \rule{0pt}{2.5ex}
			2 & \xmm & 0740920301 & 2014-12-31 &  17 \\
			& \nustar & 60001149004 &  & 22\\ \hline \rule{0pt}{2.5ex}
			3 & \xmm & 0740920401 &    2015-01-02  & 17\\
			& \nustar & 60001149006 &  & 21 \\ \hline \rule{0pt}{2.5ex}
			4 & \xmm & 0740920501& 2015-01-04  & 15\\ 
			& \nustar & 60001149008 &   & 23 \\ \hline \rule{0pt}{2.5ex}
			5 & \xmm & 0740920601 & 2015-01-06  & 21 \\
			& \nustar & 60001149010 &   & 21\\ \hline
		\end{tabular}
	\end{center}
\end{table} 
\subsection{Timing analysis}\label{subsec:lc}
\begin{figure*}
	\includegraphics[width=\linewidth]{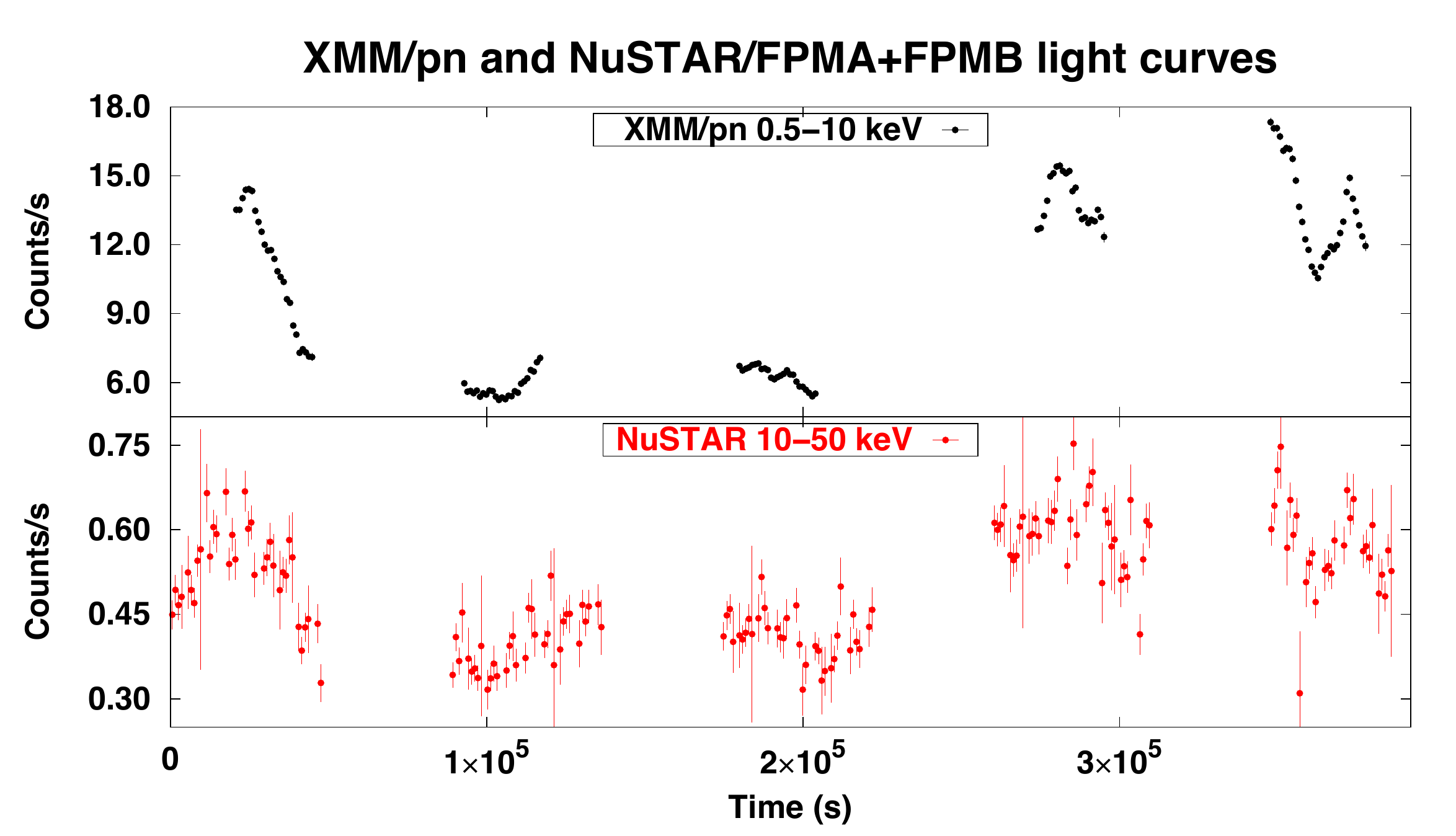}
	\caption{Count-rate light curves of the five joint \xmm~and \nustar~observations of \qc. Bins of 1 ks are used. Top panel: \xmm/pn light curves in the 0.5--10 keV band. Bottom panel: \nustar~light curves (co-adding FPMA and FPMB data) in the 10--50 keV band.
		\label{lcurves}}
\end{figure*}

\begin{figure*}
	\includegraphics[width=\linewidth]{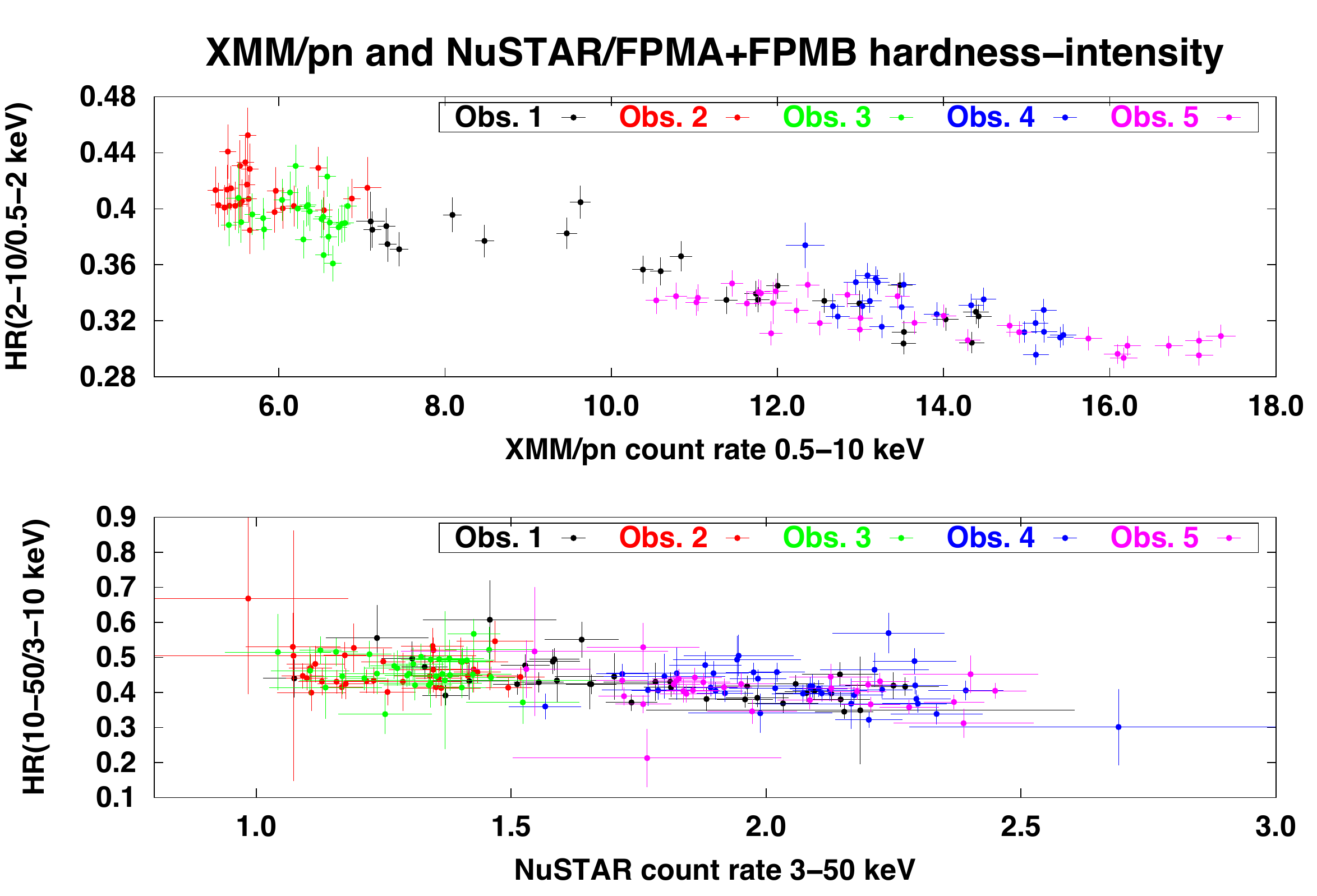}
	\caption{ ``Hardness-intensity diagram'' of the five joint \xmm~and \nustar~observations of \qc. Each point corresponds to a time bin of 1 ks. Top panel: \xmm/pn hardness ratio (2--10 keV)/(0.5--2 keV) plotted against the 0.5--10 keV count rate. Bottom panel: \nustar/FPMA+FPMB hardness ratio (10--50 keV)/(3--10 keV) plotted against the 3--50 keV count rate.
		\label{hr}}
\end{figure*}

\begin{figure*}
	\includegraphics[width=\linewidth]{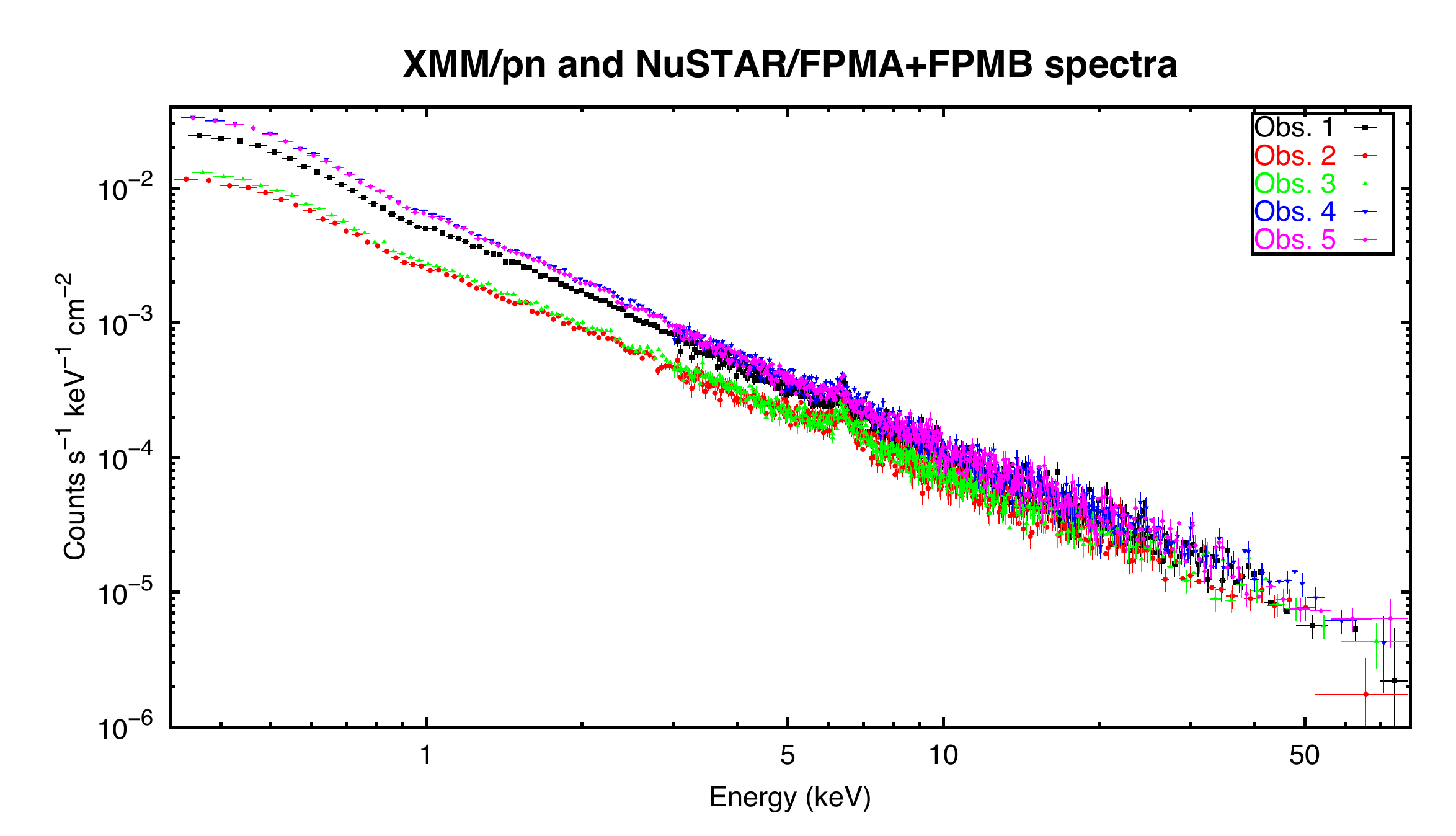}
	\caption{ The \xmm/pn and \nustar~spectra of the five observations of \qc. \nustar/FPMA and FPMB data have been co-added for plotting purposes.
		\label{ldata}}
\end{figure*}

\begin{figure*}
	\includegraphics[width=\linewidth]{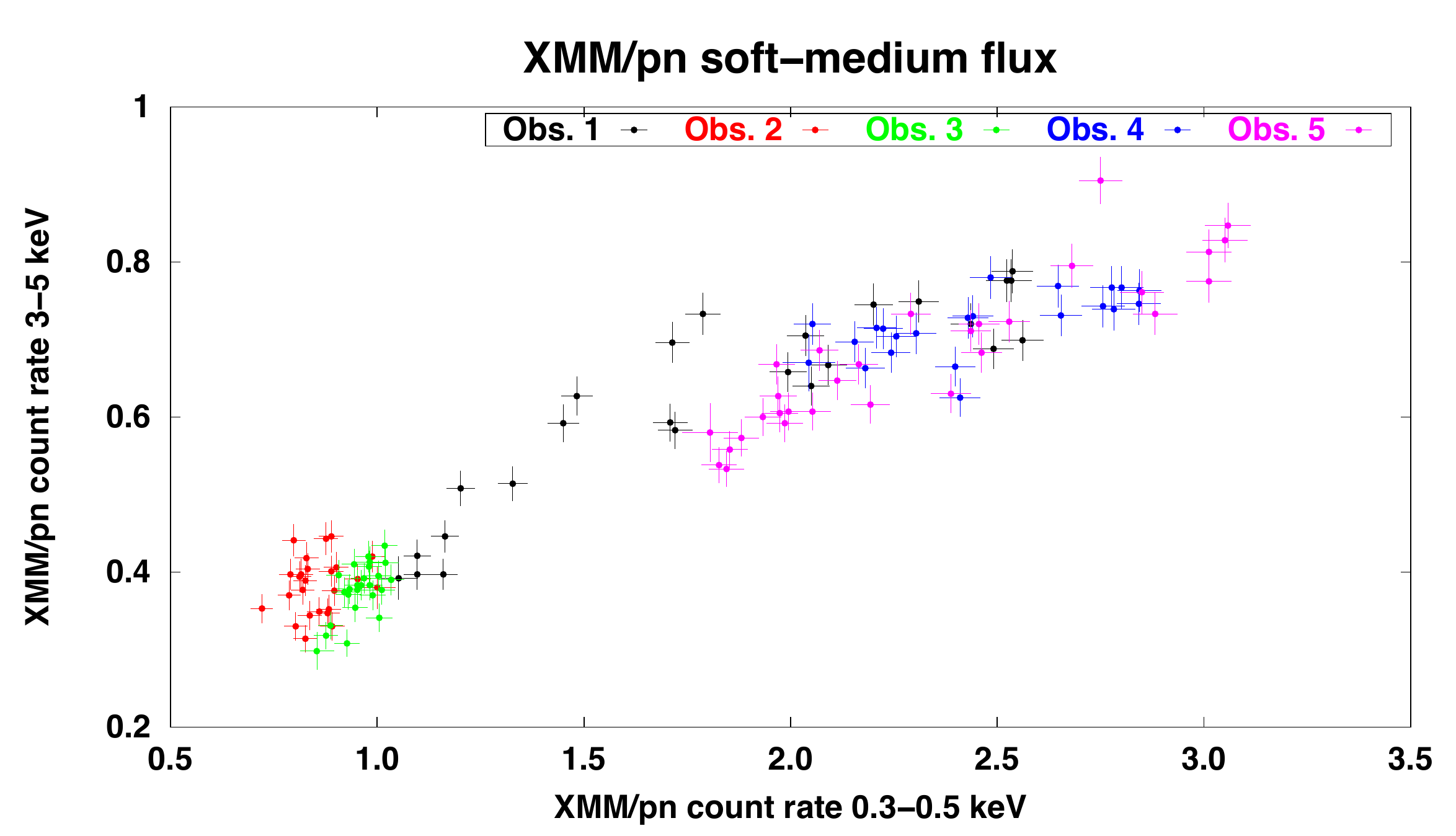}
	\caption{ \xmm/pn count rate in the 3--5 keV range plotted against the \xmm/pn count rate in the 0.3--0.5 keV range. Each point corresponds to a time bin of 1 ks.
		\label{fluxes}}
\end{figure*}

The light curves of the five observations are shown in Fig. \ref{lcurves}. Strong flux variability, particularly in the soft band, is found on time scales of a few ks. In Fig. \ref{hr}, we show the \xmm/pn (2--10 keV)/(0.5--2 keV) hardness ratio plotted against the \xmm/pn count rate in the 0.5--10 keV band, and the \nustar~(10--50 keV)/(3--10 keV) hardness ratio plotted against the \nustar~count rate in the 3--50 keV band. These plots show that a higher flux corresponds to a lower hardness ratio, i.e. the typical ``softer when brighter'' behaviour. Furthermore, the spectral variability is especially prominent in the soft band. This is also apparent from the \xmm/pn and \nustar~spectra shown in Fig. \ref{ldata}. The detailed spectral analysis is beyond the scope of this article, and is deferred to a future work (Ursini et al., in prep.).\\ \\
In Fig. \ref{fluxes} we plot the \xmm/pn count rate in the 3--5 keV range against the \xmm/pn count rate in the 0.3--0.5 keV range. The 0.3--0.5 keV band is likely dominated by the soft excess \citep{brenneman2007,markowitz2009}, while the 3--5 keV band is dominated by the primary power law-like emission with no significant contribution from the reflection component. The plot shows a clear linear relationship between the two energy bands.
\section{Discussion and conclusions}\label{sec:discuss}
We reported the first results from five joint \xmm~and \nustar~observations of \qc. The high-quality, broad-band data clearly show both flux and spectral variability on time scales as short as a few ks. The spectral variability is particularly prominent in the soft band, namely below 10 keV. Since the spectra do not show signatures of strong absorption, the observed variability appears to be intrinsic. The analysis of the hardness ratio provides a model-independent evidence that the spectrum is softer when the flux is higher, which corresponds to the ``softer when brighter'' behaviour generally found in luminous Seyfert galaxies \citep[see, e.g.,][]{sobolewska&papadakis,soldi2014}. This behaviour can be understood as an effect of Comptonization, since an increase of the soft seed photon luminosity, due for example to a higher accretion rate, results in a more efficient cooling of the hot corona via inverse Compton scattering. A lower coronal temperature in turn determines the steepening of the spectrum. \\ \\ 
The data show a linear relationship between the fluxes in the 3--5 keV energy band and in the 0.3--0.5 keV band, where the expected dominant components are the primary power law and the soft excess, respectively. This linear relationship suggests a common physical origin for the hard X-ray emission and the soft excess. If the hard X-ray emission is due to Comptonization, this result is consistent with the soft excess being produced via Comptonization of the same seed photons \citep[see also][]{brenneman2007,markowitz2009}. This may in turn suggest the presence of a both a ``hot'' corona with a temperature of 10--100 keV, which is responsible for the hard X-ray emission, and a ``warm'' corona with a lower temperature (e.g. $\sim 0.5$ keV), which is responsible for the soft excess \citep[for a detailed discussion of this scenario, see][]{pop2013mrk509}. In this case, geometrical or physical variations of the accretion disc/corona system, such as changes in the accretion rate, could account for both the soft excess and primary continuum variability.\\ \\
A detailed spectral analysis, with the study of the soft excess and the reflection component, will be  presented in a forthcoming paper, also including the UV data from the \xmm~Optical Monitor, which may carry useful informations to study the Comptonization components \citep[see, e.g.,][]{pop2013mrk509}. 
We stress that this is the first \xmm+\nustar~monitoring program explicitly aimed at the study of the high-energy emission of an AGN through its variability on time scales of days down to a few ks. In the framework of a previous campaign on \cc, conducted over a few months, the broad-band coverage of \textit{XMM-Newton, NuSTAR} and \textit{INTEGRAL} allowed us to study the high-energy spectrum of that source in detail \citep{5548}. The approach based on multiple, broad-band observations appears to be very promising for a better understanding of the processes associated with the high-energy emission of AGNs. 
Besides, further studies on a greater number of sources will be needed to investigate the similarities and differences between different types of AGNs.
\section*{Acknowledgements}
We are grateful to the referee, Erin Kara, for comments that improved the paper.\\
This work is based on observations obtained with: the \nustar~mission, a project led by the California Institute of Technology, managed by the Jet Propulsion Laboratory and funded by NASA; \xmm, an ESA science mission with instruments and contributions directly funded by ESA Member States and the USA (NASA). This research has made use of data, software and/or web tools obtained from NASA's High Energy Astrophysics Science Archive Research Center (HEASARC), a service of Goddard Space Flight Center and the Smithsonian Astrophysical Observatory. 
FU, POP, GM, SB, MC, ADR and JM 
acknowledge support from the french-italian International Project of Scientific Collaboration: PICS-INAF project number 181542. FU, POP acknowledge support from CNES. FU acknowledges support from Universit\'e Franco-Italienne (Vinci PhD fellowship). FU, GM acknowledges financial support from the Italian Space Agency under grant ASI/INAF I/037/12/0-011/13. SB and MC acknowledges financial support from the Italian Space Agency under grant ASI-INAF I/037/12/P1. 
GP acknowledges the Bundesministerium f\"ur Wirtschaft und Technologie/Deutsches Zentrum f\"r Luft- und Raumfahrt (BMWI/DLR, FKZ 50 OR 1408) and the Max Planck Society for support. 
\bibliographystyle{aa}
\bibliography{mybib.bib}
\end{document}